\documentclass[aps,prd,twocolumn,superscriptaddress,nofootinbib]{revtex4-2}

\usepackage{graphicx}
\usepackage{amsmath,amssymb}
\usepackage{booktabs}
\usepackage[colorlinks=true,linkcolor=blue,citecolor=blue]{hyperref}
\usepackage[usenames,dvipsnames]{xcolor}
\usepackage{enumitem}

\definecolor{claudeorange}{rgb}{0.85,0.42,0.06}

\newcommand{\mjj}{m_{jj}}
\newcommand{\pt}{p_{\mathrm{T}}}
\newcommand{\epsc}{\epsilon_c}
\newcommand{\epsb}{\epsilon_b}
\newcommand{\epsl}{\epsilon_\ell}
\newcommand{\TeV}{\ensuremath{\,\mathrm{TeV}}}
\newcommand{\GeV}{\ensuremath{\,\mathrm{GeV}}}

\setlist[itemize]{leftmargin=1.2em, itemsep=1pt, topsep=2pt, parsep=0pt}

\begin{document}

\title{The era of charmed resonance searches at the LHC}

\author{Daniel Whiteson}
\affiliation{Department of Physics \& Astronomy, University of California, Irvine}

\begin{abstract}
The LHC dijet program has searched extensively for resonances decaying to inclusive jets and, more recently, to bottom-quark jets, where the sensitivity enhancement from suppression of the background outweighs the loss of signal efficiency in bottom-quark tagging.  However, no dedicated search for dijet resonances containing charm-tagged jets has been performed. We study the charm-tagging performance required to make such searches competitive with the inclusive dijet search. Using the ATLAS 13 TeV dijet mass spectrum with 139 fb$^{-1}$, we predict the expected charm-tagged spectra from simulated jet-flavor fractions and parameterized charm-, bottom-, and light-jet tagging efficiencies, and estimate the sensitivity of a profile-likelihood search in exclusive one- and two-tag categories. At the loosest working point of the current tagger, charm-tagged searches improve the expected sensitivity to a resonance decaying to two charm quarks across the full resonance mass range considered, although the quantitative sensitivity depends on the poorly constrained high-$\pt$ behavior of charm-tagging efficiencies and on the simulated heavy-flavor composition of the dijet background. These results provide performance targets for charm taggers and motivate the development of dedicated charm-tagged dijet resonance searches at the LHC.

\end{abstract}

\maketitle

% =====================================================================
\section{Introduction}

Searches for resonances in the dijet spectrum have broad discovery power for particles coupling to quarks or gluons~\cite{HarrisKousouris,ATLASdijet2019,CMSdijet2019} and are flagship elements of the physics program at the Large Hadron Collider.  The ATLAS and CMS collaborations have searched the flavor-inclusive spectrum through the full Run~2
        \cite{ATLASdijet2019,CMSdijet2019} dataset. Major challenges of the dijet search program include characterizing and reducing the copious QCD-dominated background spectrum.

        One fruitful approach is to search for resonances decaying predominantly to specific quarks, e.g. heavy flavor bottom ($b$) quarks, where the reduced background levels justify the efficiency cost of $b$-tagging the signal~\cite{ATLASdijet2019,ATLASbdijet2018,CMSbdijet2018}. To date, this strategy has not been generalized to charm quarks, and there have been no dedicated dijet searches making use of $c$-tagging algorithms.  Historically, charm identification was too inefficient, and the charm content of the QCD background too large to make such a search competitive with the flavor-blind analysis, even for resonances decaying to charm pairs.
        
    However, dedicated charm taggers have since improved substantially~\cite{ATLASGN2,ATLASGN3,CMSBTV20001,CMS:2025kje}, recently enabling deployment on a resonance search to $\gamma+$jet~\cite{ATLASgammajet2026}. Modern $VH(H\rightarrow c\bar{c})$ searches~\cite{ATLASVHcc2025,CMSVHcc2023,CMSttHcc2025,ATLAScHgg2025} demonstrate  the substantial gains achievable with improved charm identification.

    This raises the questions: what level of performance is necessary to open the era of charmed dijet resonance searches? Are the current charm taggers sufficient to allow for a dedicated search for dijet resonances in the charm-tagged sample? These questions are unanswered in the literature.

    In this paper, we (i) model current charm-tagging performance in a dijet resonance search to predict the charm-tagged background, (ii) explore the sensitivity of LHC datasets to resonances decaying to charm pairs or charm$+$gluon\footnote{ Such flavor-nonuniversal couplings are constrained by flavor observables, but models with enhanced couplings to charm remain viable~\cite{PhysRevD.74.013011,Arhrib:2006sg,Hohne:2023uel}.}, and (iii) explore the performance  required to allow for dedicated charm-tagged dijet searches.

    Section~\ref{sec:appr} gives an overview of our approach. In Sec.~\ref{sec:flav} we describe our modeling of flavor fractions and jet $\pt$ distributions in the dijet spectrum, as well as the associated uncertainties. 
    Section~\ref{sec:charm} applies the model to charm tagging and predicts the sensitivity of a dedicated search in the charm-tagged dijet spectrum.

% =====================================================================
\section{Approach}
\label{sec:appr}

We estimate the sensitivity of the LHC dataset to two classes of resonances: a $Z'$-like vector boson which decays to a pair of charm quarks, and a vector-like quark $T$ which decays to $cg$~\cite{Tong:2023VLQ}. The sensitivity is driven by the size of the charm-tagged background, and its uncertainty.

We predict the charm-tagged background by applying a tagging model to the ATLAS published untagged dijet spectrum from a sample of $pp$ collisions at $\sqrt{s}=13$ TeV with integrated luminosity of 139 fb$^{-1}$~\cite{ATLASdijet2019}.  Numerically the number $N$ of tagged  events in a bin centered at $\mjj$
\begin{equation}
          N_{\rm tag}(\mjj) = N_{\rm untag}(\mjj)\;A(\mjj)\;\kappa(\mjj)
        \end{equation}

\noindent
depends on $A(m_{jj}) \equiv A_{\mathrm{tagged}}(m_{jj})/A_{\mathrm{untagged}}(m_{jj})$,
the ratio of the fiducial acceptances of the tagged and untagged selections, and $\kappa$, the fraction of events in the tagged fiducial region which pass the tag, both of which are measured in simulated samples.

Predicting the charm-tagged fraction $\kappa$ requires (i) a model of the flavor fractions of the QCD background as a function of jet $\pt$, (ii)  the tagging efficiencies  $\epsilon_\ell, \epsilon_c,\epsilon_b$ for light ($\ell)$, charm ($c$) and bottom quarks ($b)$ as a function of jet $\pt$, and (iii) the distribution of jet $\pt$ within each $\mjj$ bin.   

Since a specific bin can receive contributions from many flavor pairs $ij \in [{bb,bc,b\ell,cc,c\ell,\ell\ell]}$ and many pairs of jet $\pt$ bins $ab$, where often $a \ne b$, we calculate the two-charm-tag fraction $\kappa_2$ as a sum over all flavor pairs and jet $\pt$ bin pairs:
        \begin{equation}
          \kappa_2(\mjj) = \frac{\sum_{ij,ab} w_{ij,ab}\,\epsilon_i(\pt^a)\,\epsilon_j(\pt^b)}
                          {\sum_{ij,ab} w_{ij,ab}}
                          \label{eq:k}
        \end{equation}

\noindent
where $w_{ij,ab}(\mjj)$ is the fraction of simulated dijet events with flavor $ij$ in jet $\pt$ bins $ab$ within dijet mass bin centered at $\mjj$.

Similarly, the dijet spectrum where at least one jet is tagged is built from

\begin{equation}
  \kappa_{\geq 1}(\mjj) = 1 -
    \frac{\sum_{ij,ab} w_{ij,ab}\,
          \left[1-\epsilon_i(\pt^a)\right]
          \left[1-\epsilon_j(\pt^b)\right]}
         {\sum_{ij,ab} w_{ij,ab}}
\end{equation}

\noindent
and the exactly-one-tag version is $\kappa_1 = \kappa_{\geq 1} - \kappa_2$.  Our estimation of the flavor fractions and efficiencies as well as their uncertainties are discussed in detail in the next section.

With a prediction of the tagged background spectrum and a model of the resonance mass shape, acceptance and efficiency, we can calculate expected limits on the product of the resonance cross section and branching fraction. 

We fit the background spectrum using the standard four-parameter dijet function~\cite{CDFdijet2009}, with all parameters free and profiled. No other background shape systematics were used.  We fit the untagged spectrum to better model the expected distribution before applying $A\kappa$, which is also smoothed to remove simulation scatter.  Limits at 95\% CL are computed using a profile likelihood ratio in the asymptotic approximation~\cite{CowanAsymptotic} with the $\mathrm{CL}_s$ test statistic~\cite{ReadCLs,JunkCLs}. 
For a signal of a narrow Gaussian, our machinery recovers the ATLAS limits on the same signal shape~\cite{HEPDATAdijet2019} to within 10\%; see Appendix~\ref{app:stats}.

Limits from the untagged and tagged spectra are computed with the identical machinery, and the crucial metric is their ratio, which determines whether the tagged spectrum's greater background reduction but lesser signal efficiency gives it improved statistical power.

% =====================================================================
\section{Jet flavor fractions and $\pt$ }
\label{sec:flav}

To predict the tagged dijet mass spectrum from the untagged spectrum requires application of the tagging efficiency to the jets.  However, that efficiency is typically strongly dependent on jet $\pt$ and flavor, and so requires an estimate of the flavor fractions and transverse momenta within each bin.  In this section, we describe the use of samples of simulated dijet events to estimate the jet $\pt$ distribution within each mass bin, and the flavor fractions as a function of jet $\pt$. Together with the $b$- or $c$-tagging efficiency models, described later, this allows for a prediction of the tagged spectrum.  

Samples of simulated dijets are used, as no measurement of the dijet flavor composition is available at
        $\sqrt{s}=13$~TeV. However, measured fractions are available at $\sqrt{s}=7$~TeV~\cite{ATLAS7TeVflavour}.  %We validate our model in the lower energy data, derive a correction, build an uncertainty band, and cross check the corrected 13~TeV composition against ATLAS        simulation~\cite{ATLASCONF2016060}.
QCD dijet events are simulated with \textsc{Pythia}~8.3~\cite{Pythia83} at $\sqrt{s}=7$ and 13 TeV, using ATLAS A14 tune~\cite{A14tune} with
        NNPDF2.3LO~\cite{NNPDF23}. Anti-$k_t$ jets~\cite{antikt,FastJet} are reconstructed with radius parameter $R=0.4$, built from stable particles. Jets are required to have $\pt>50$ GeV and $|\eta|<2.8$ and $|y^*|<0.6$  in the untagged sample and $|\eta|<2.0$ and $|y^*|<0.8$ in the tagged, following the ATLAS $b$-tagged search.
        
        A jet is labelled $b$ if a weakly decaying $b$-hadron with
        $p_\mathrm{T}>5$~GeV lies within $\Delta R < 0.3$ of the jet axis;
        otherwise $c$ under the same requirement; otherwise light. This definition is used to allow direct comparison with ATLAS results.  We measure the joint dijet composition rather than the single-jet, as a tag may be applied to either jet.  For each dijet mass bin, and each flavor pair, events are sorted into a two-dimensional histogram in jet $\pt$ with $8\times 8$ bins, which allows calculation of the $w_{ij,ab}$ factor  in Eq.~\ref{eq:k}.  

        We validate our model in the lower energy data, where it describes the
measured fractions well except in $f_{b\ell}$; we correct for this with a
shift $\delta = +0.0073 \pm 0.0024$ in absolute event fraction from $c\ell$
to $b\ell$, build an uncertainty band, and cross check the corrected 13~TeV
composition against ATLAS simulation~\cite{ATLASCONF2016060}. See
App.~\ref{app:flavor} for details.

\begin{figure}[ht]
  \centering
  \includegraphics[width=0.90\columnwidth]{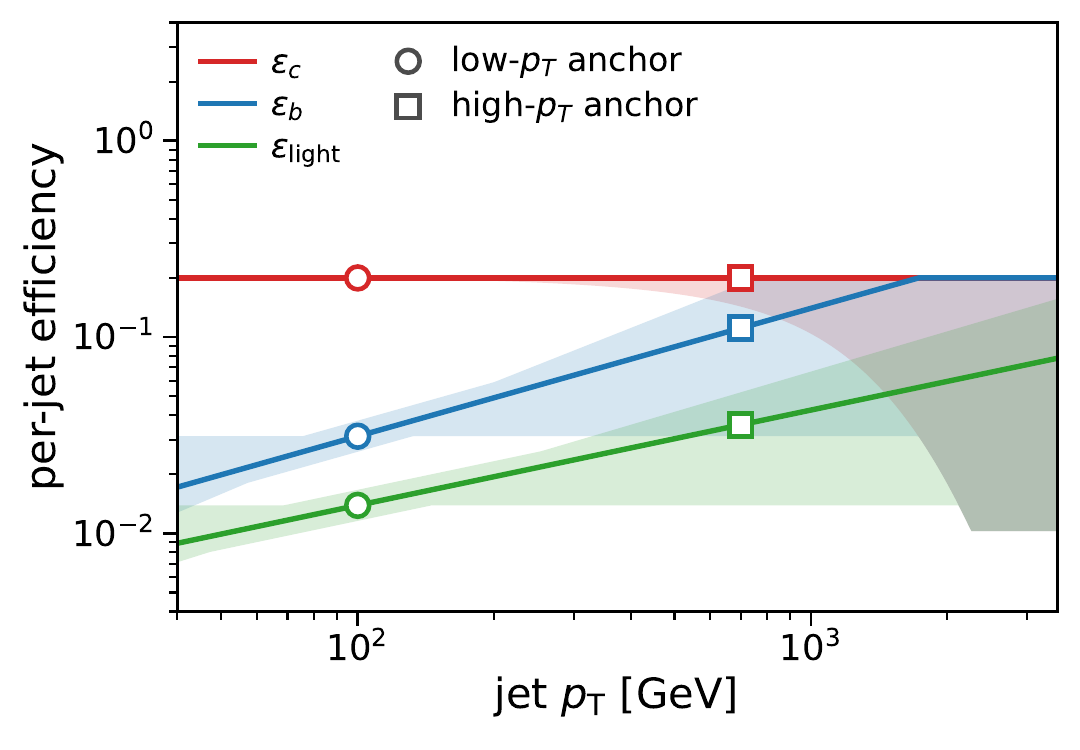}
    \includegraphics[width=0.90\columnwidth]{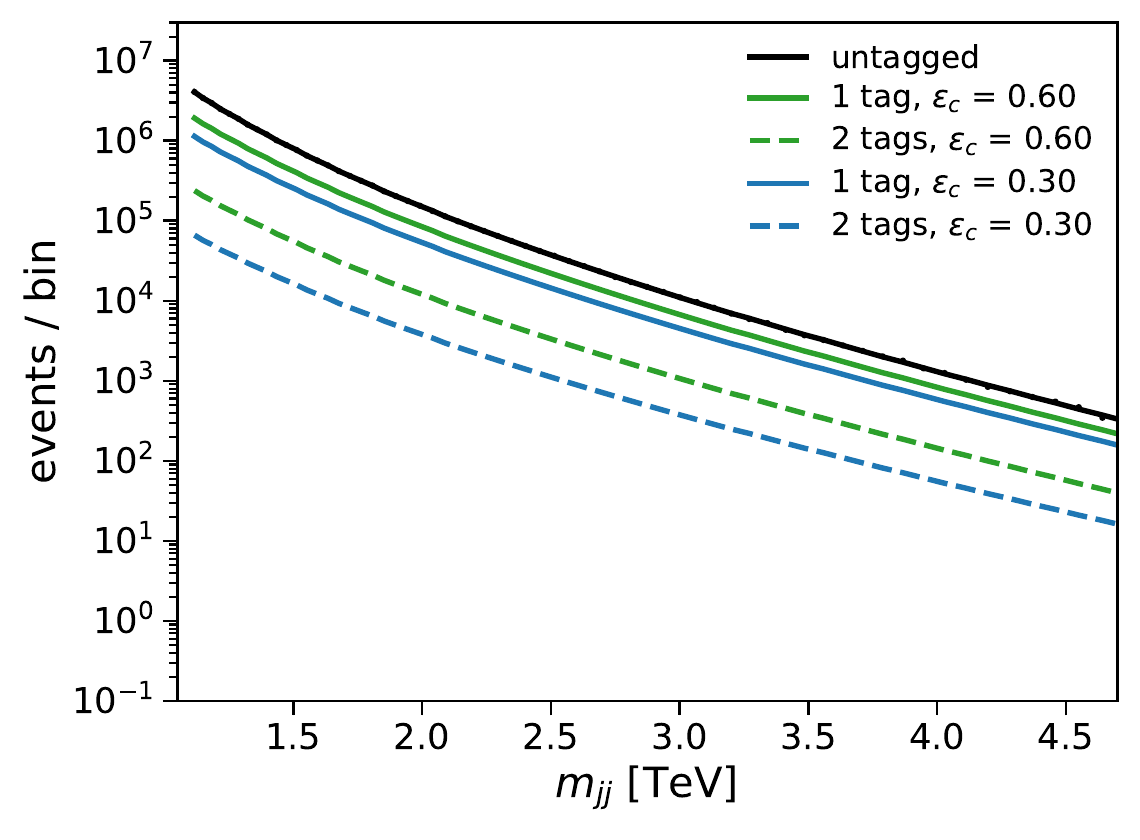}

  \caption{Top: published charm-tagging  efficiency for charm-,bottom- and light-quark jets in two samples spanning
  different jet $\pt$ ranges, and our interpolation. Bottom: observed untagged dijet spectrum~\cite{ATLASdijet2019} and the predicted 1- or 2-charm-tag
  spectra at two working points using our method in the text.}
  \label{fig:charmeff}
\end{figure}

% =====================================================================
\section{Application to charm tagging}
\label{sec:charm}

With a model of the jet flavor and momenta in each bin, predictions of the tagged spectrum require a description of the tagging performance. We validate the procedure and the flavor model by applying it to $b$-tagging, where the observed tagged spectra are available for comparison; see App.~\ref{app:btag} for details. 

To predict the charm-tagged spectra, we apply the efficiencies of the GN2 charm single-jet tagger~\cite{ATLASGN2}, a relatively recent and well-documented algorithm. For simplicity, we use the same fiducial definition for the tagged region as ATLAS $b$-tagged dijet search, relative to the untagged, and do not optimize further. The tagger offers several charm-efficiency working points.

While several efficiencies are published, explicit dependence of the tagger performance as a function of jet $\pt$ is not available. However, efficiencies for two samples with distinct $\pt$ ranges anchor the model and demonstrate the rough dependence on jet $\pt$.

We hold $\epsc$ at the chosen working point\footnote{Fixing $\epsc$ corresponds to a working point recalibrated in
each $\pt$ bin.}
and let the mis-tag rates $\epsb$ and $\epsl$ rise with $\pt$, interpolating
log-linearly between the two published values for each working point and continuing that trend
beyond them; see Fig.~\ref{fig:charmeff};      several  variations on these assumptions are also explored in Appendix~\ref{app:scenarios}, and do not qualitatively impact the conclusions. 

In the $b$-tagging study of App.~\ref{app:btag}, we see a slight preference for an enhanced tagging efficiency for jets with multiple heavy hadrons, the fit prefers $\alpha=1.96$ for this flavor and tagging model; see Eq.~\ref{eq:alpha}. In the charm-tagging case,  we similarly apply an efficiency which depends on the number of hadrons found within the jet, using the same value of $\alpha$, which assumes similarity in the {\it dependence} of the $b$- and $c$-tagging on the number of hadrons, rather than the prevalence of multi-hadron jets or the efficiencies themselves.  
Because a larger $\alpha$ raises the efficiency of the multi-hadron jets that populate the background while leaving the predominantly single-hadron signal unaffected, $\alpha = 1.96$ corresponds to the conservative end of the range; $\alpha$ is varied down to 1.0 which also accommodates the data and corresponds to independent tagging opportunities for each heavy hadron, to bound the sensitivity of the projections to this extrapolation. The fraction of jets with multiple hadrons, $f_{\rm{multi}}$, in the $Z'\rightarrow c\bar{c}$ signal is small: $4\%-6\%$, rising with mass.

 The charm tagged spectra at two working points are shown in Fig.~\ref{fig:charmeff} using two exclusive categories, $=1$ or $=2$ tags.

We estimate the sensitivity of a search for two classes of resonances in the charm-tagged spectrum.  The first is a massive vector boson, a charmophilic $Z'$, which is generated in Pythia with the same showering and reconstruction as the background sample. The second resonance is a vector-like quark $T$ with charge $2/3$, produced singly and decaying
        to charm plus a gluon~\cite{Tong:2023VLQ}. The coupling is through a dimension-five dipole operator:
        \begin{equation}
          \mathcal{L} \supset
            \frac{g^i_g}{\Lambda}\,\bar T \sigma^{\mu\nu} u_{R,i}\,G_{\mu\nu}
          + \frac{g^i_\gamma}{\Lambda}\,\bar T \sigma^{\mu\nu} u_{R,i}\,F_{\mu\nu}
        \end{equation}
        which couples only to charm when $g^1 = g^3 = 0$, $g^2 = 0.1$,  has a mass-independent branching ratio to $cg$ of 0.4 and a narrow width across the mass range. Samples of simulated  $T\rightarrow cg$ events are generated  MadGraph5~\cite{MadGraph5aMCNLO} 3.5.11, showered and
        hadronized with the same configuration as the $Z'$ benchmark.

        Flavor limits are avoided by choosing       $\Lambda = 3\TeV$~\cite{Tong:2023VLQ}. The $T$ also decays to $c\gamma$, which is already excluded~\cite{ATLASgammajet2026}, but no limits exist on the $cg$ channel.  
        Flavor-nonuniversal couplings are not generic in all extensions of the Standard Model and can be constrained by flavor observables, particularly when they induce flavor-changing interactions. Nevertheless, models with enhanced couplings to particular quark flavors remain viable, and the experimental sensitivity to such scenarios is strongly dependent on the available flavor-tagging performance. We  do not attempt to motivate a particular flavor model; instead, we determine the charm-tagging performance required to improve upon an inclusive dijet search. 
        
\begin{figure}[t]
  \centering
    \includegraphics[width=\columnwidth]{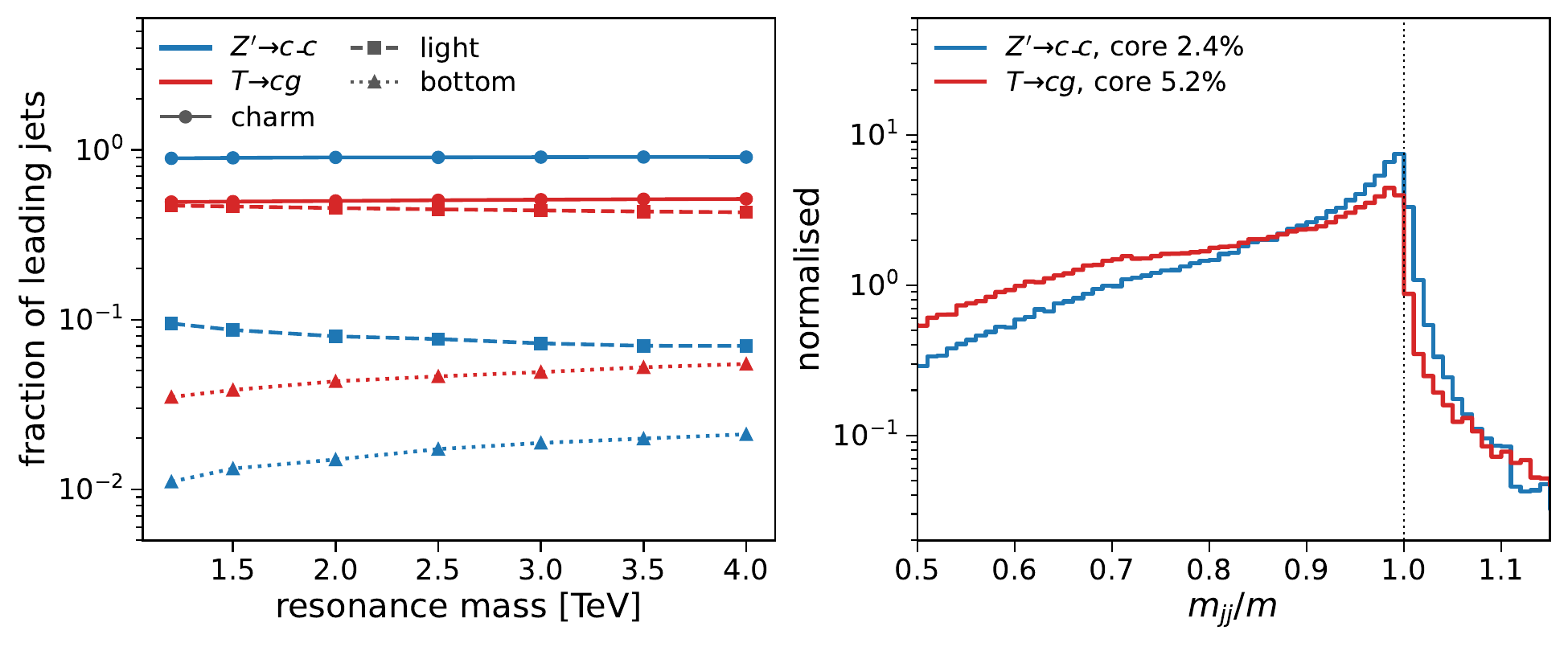}
  \caption{Properties of the $Z'\to c\bar c$ and $T\rightarrow cg$ resonances. Left, the flavor fraction of leading jets versus resonance mass. Right, the distribution of invariant masses from two leading jets, $m_{jj}$ relative to the true resonance mass.}
  \label{fig:signal}
\end{figure}

        To estimate the sensitivity, we measure the flavor composition of the $Z'$ or $T$ decay jets, the reconstructed $\mjj$ distribution and the signal acceptance ratio between the untagged and tagged fiducials. The flavor composition and mass distributions are shown in Fig.~\ref{fig:signal}.  The acceptance is larger in the tagged fiducial region, with
$A_{c\text{-tagged}}/A_{\mathrm{untagged}} = 1.19\text{--}1.27$ for
$Z' \to c\bar{c}$ and $1.19\text{--}1.22$ for $T \to cg$ over the mass range
considered.
%in both cases well below the background ratio 
%since a resonance produces central jets that largely pass the tighter $|y^*|$
%requirement.

An event falls into either the $=1$ or $=2$ tag categories, and never both, so they
        are statistically independent and the likelihood is a simple product
        with no correlation term:

   \begin{align}
  L(\mu,\vec p_1,\vec p_2) =\;
    & \prod_{1k} \mathrm{Pois}\!\left[n_{1k} \mid \mu s_{1k}
                                    + b_{1k}(\vec p_1)\right] \nonumber \\
    \times\; & \prod_{2k} \mathrm{Pois}\!\left[n_{2k} \mid \mu s_{2k}
                                    + b_{2k}(\vec p_2)\right]
\end{align}

This likelihood has one shared signal strength $\mu$, with two independent four-parameter background
        shapes, for a total of eight profiled nuisance parameters.  
 The single-tag category has more signal but worse purity, while the
        two-tag has the reverse properties, making the combination of the exclusive categories  more powerful than the single $\ge 1$ inclusive or the $=2$ category individually.
        
We calculate the expected limits as described above, for both the tagged and untagged cases.  The figure of merit is $G$, the relative gain in limits on the product of the cross section and branching ratio, where

\[ \sigma^{95} = \frac{N^{95}}{\mathcal{L} A \epsilon} \]

\noindent
and the gain is therefore

\[
  G(m) = \frac{\sigma^{95}_{\rm untagged}(m)}{\sigma^{95}_{c-{\rm tagged}}(m)}
       = \frac{N^{95}_{\rm untagged}}{N^{95}_{c-{\rm tagged}}}\;
         \frac{A_{c-{\rm tagged}}}{A_{\rm untagged}}\;\epsilon_{c-{\rm tag}}
\]

\noindent
where $\epsilon_{c-{\rm tag}}$ is the probability that a signal event in the
tagged fiducial region is retained by the tag, summed over the $=1$ and $=2$
tag categories, and $\epsilon_{\rm untagged}=1$.  The event-level tagging probability $\epsilon_{c-{\rm tag}}$ is calculated in the same way
as $\kappa$ in Eq.~\ref{eq:k}, but from the flavor composition of the signal
jets.  A value of $G>1$ indicates a gain, where the improved (smaller) limit at 95\% CL on the number of signal events $N^{95}$ overcomes the cost of the tagging efficiency.   The branching fraction and the luminosity cancel, since the same signal appears in both limits, so $G$ can be expressed as a ratio of cross sections.  

\begin{figure}[t]
  \centering
  \includegraphics[width=\columnwidth]{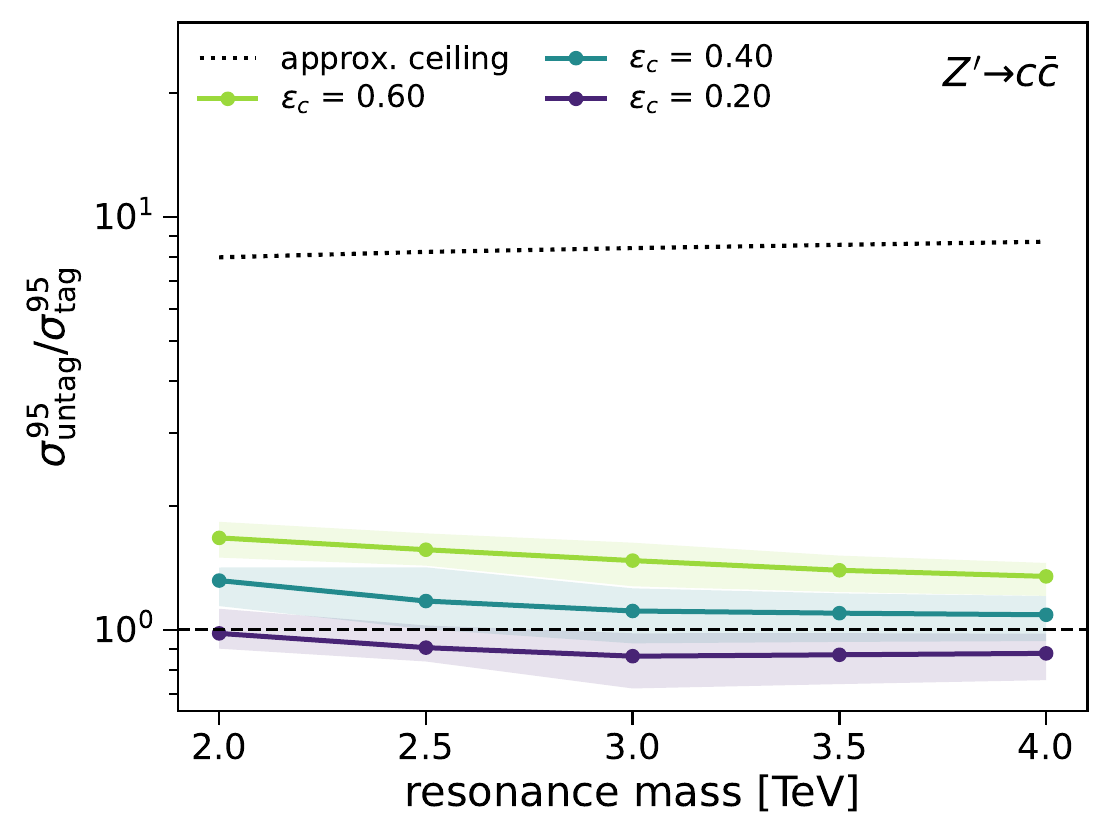}
\includegraphics[width=\columnwidth]{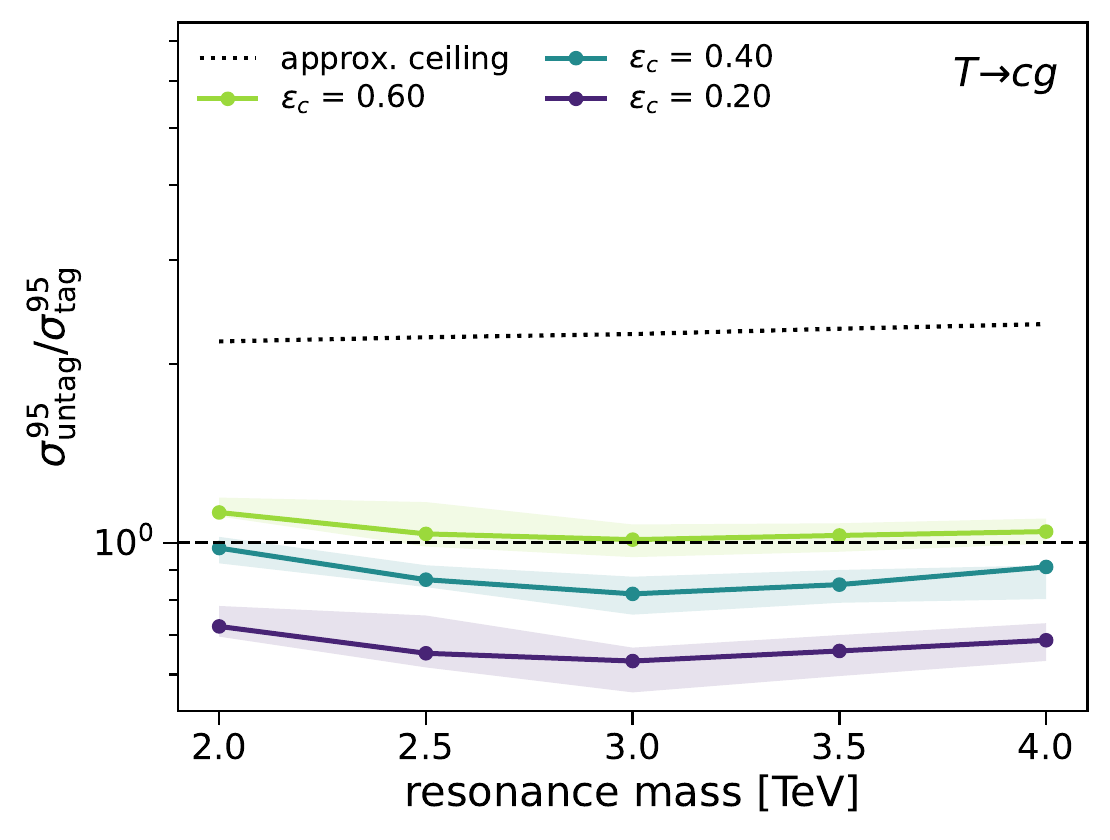}
  \caption{ The relative gain in sensitivity of a tagged search to an untagged search. A ratio greater than one indicates the tagged search is more powerful (smaller $\sigma$), shown  versus resonance mass for several working points. Also shown is 
  the approximate ceiling, the gain a perfect cone-label tagger would
  reach against the same background.  Bands span the variations in the charm-efficiency model as well as flavor and other uncertainties.   Top, $Z'\rightarrow c\bar{c}$; bottom, $T\rightarrow cg$.}
  \label{fig:limits}
\end{figure}

Figure~\ref{fig:limits} shows the relative gain $G$ versus mass for several working points, for $Z'$ and $T$ signals.  For the $Z'$, the $\epsilon_c=0.6$ working point has a gain greater than 1 across the full mass range; for the $T$, there is gain for lower masses. A charm-tagged spectrum gives greater sensitivity to these resonances than the untagged, even for the resonance with only a single charm leg.

The band in the figures represents variations of the assumptions made in our projections, such as charm-tagging efficiency or showering at 13 TeV; see Table \ref{tab:band_zpcc}.  These are not treated as nuisance parameters of the search, because they reflect information we are lacking but will likely be available to the collaborations. The dominant effect is uncertainty in the charm-tagging efficiency.

\begin{table}[htbp]
  \centering
  \caption{Contributions to the width of the sensitivity band on $G$ for $Z' \to c\bar{c}$ at working point $\epsilon_c=0.6$.  Shown is the percent variation of $G$ that captures the envelope of variations within that source, see text for details.  The total is the sum of each source in quadrature. Values for  $T\rightarrow c g$ are qualitatively similar, dominated by tagging uncertainties.}
    \label{tab:band_zpcc}
  \begin{tabular}{lccccc}
    \toprule
     & \multicolumn{5}{c}{$m$ [TeV]} \\
    \cmidrule(lr){2-6}
     Source & 2.0 & 2.5 & 3.0 & 3.5 & 4.0 \\
    \midrule
    Charm-tagging model  & $^{+7.7}_{-9.5}$ & $^{+8.4}_{-7.8}$ & $^{+9.3}_{-12.5}$ & $^{+7.6}_{-10.8}$ & $^{+7.0}_{-9.7}$ \\
    FSR shower scale  & $^{+3.8}_{-4.3}$ & $^{+3.0}_{-3.2}$ & $^{+3.3}_{-4.6}$ & $^{+2.7}_{-3.8}$ & $^{+2.5}_{-3.3}$ \\
    PDF set  & $^{+2.5}_{-0.0}$ & $^{+2.6}_{-0.6}$ & $^{+2.5}_{-0.1}$ & $^{+2.0}_{-0.1}$ & $^{+1.8}_{-0.0}$ \\
    $b$-light corr. $\delta$ & $^{+0.3}_{-0.1}$ & $^{+0.7}_{-0.5}$ & $^{+0.1}_{-0.1}$ & $^{+0.0}_{-0.1}$ & $^{+0.1}_{-0.0}$ \\
    Multi-hadron tagging  & $^{+2.7}_{-0.0}$ & $^{+2.1}_{-0.0}$ & $^{+2.7}_{-0.0}$ & $^{+1.9}_{-0.0}$ & $^{+1.7}_{-0.0}$ \\
    \cmidrule(lr){1-6}
    \textbf{Total } & $^{+9.4}_{-10.4}$ & $^{+9.5}_{-8.5}$ & $^{+10.5}_{-13.3}$ & $^{+8.5}_{-11.4}$ & $^{+7.8}_{-10.3}$ \\
    \bottomrule
  \end{tabular}
\end{table}

The $Z'$ resonance benefits from the additional charm, where the fraction of leading jets that are genuinely charm is roughly twice as high. In addition, the gluon jet from the $T$ has a broader radiation pattern, with more energy falling outside of the cone, which translates into a longer low-mass tail and weaker sensitivity.

\begin{figure}[t]
  \centering
  \includegraphics[width=\columnwidth]{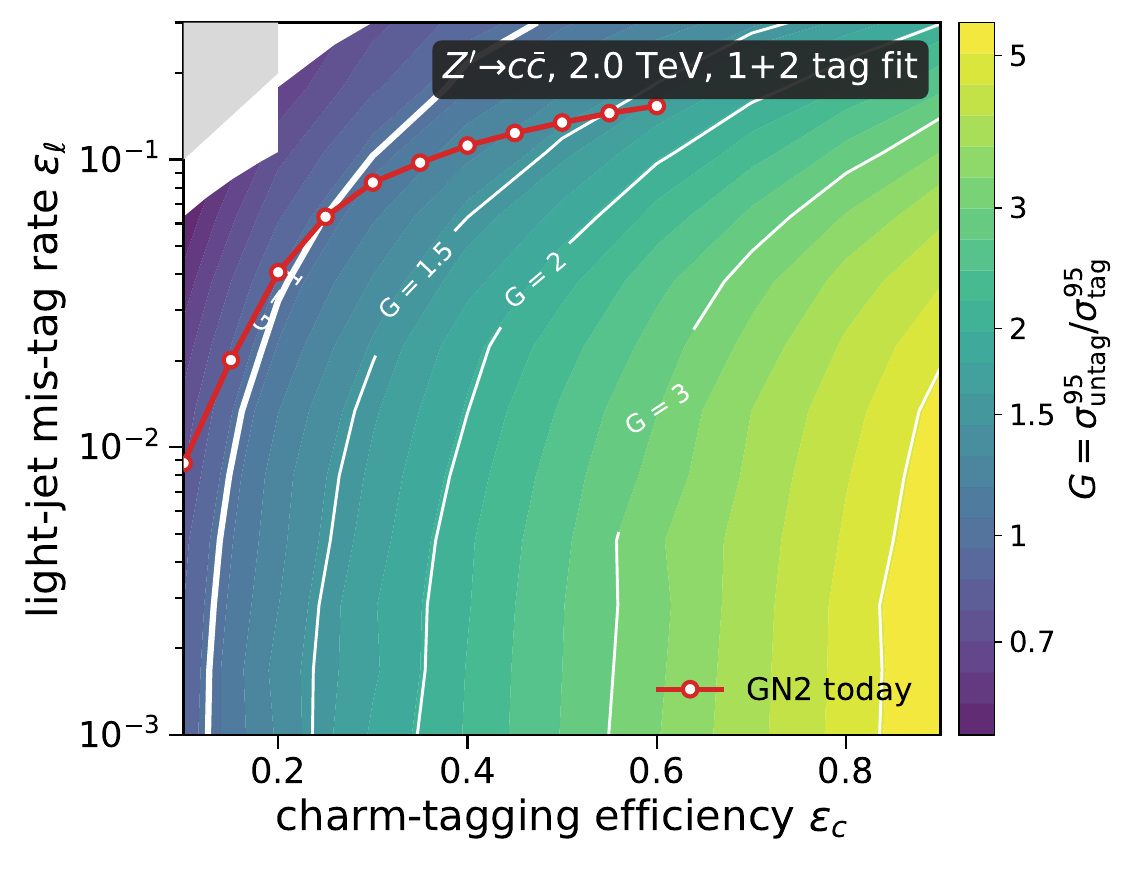}
  \includegraphics[width=\columnwidth]{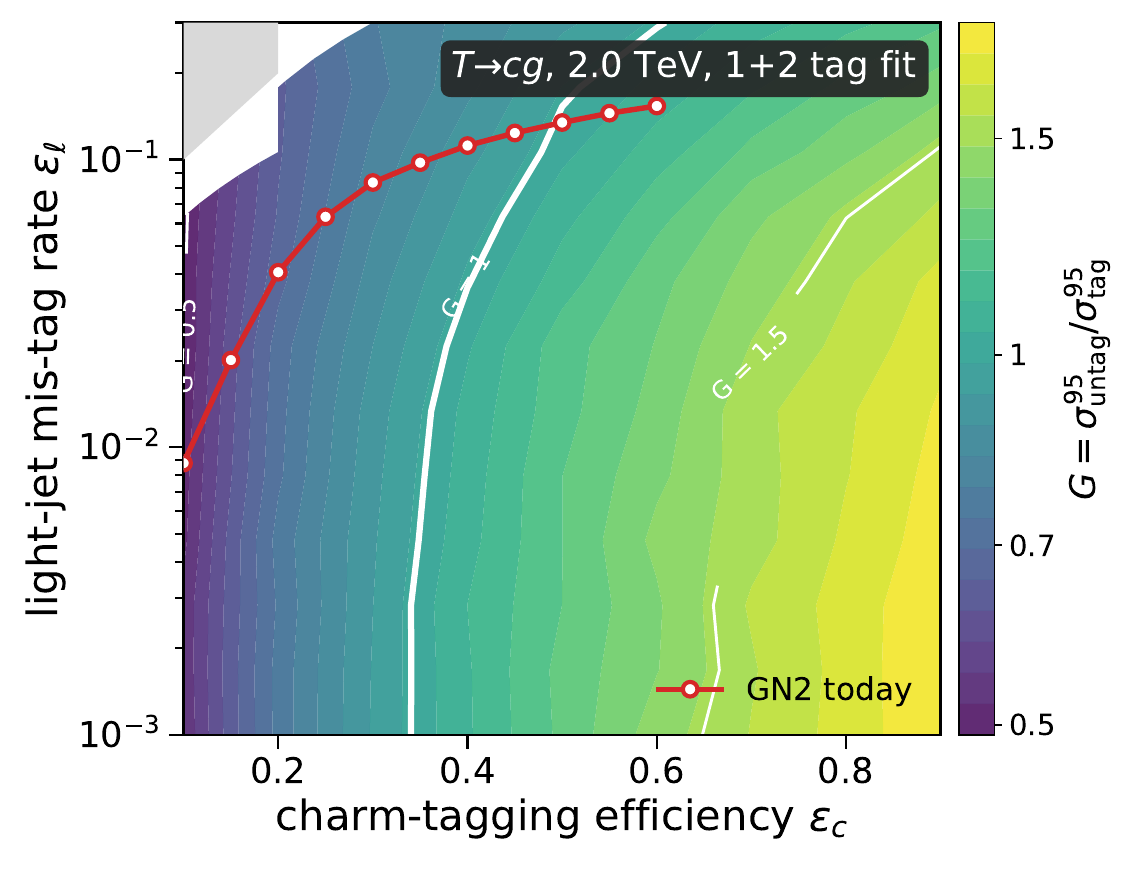}
  \caption{Relative gain $G$ over the plane of charm efficiency and light-jet mis-tag rate, at a resonance mass of $2\TeV$ for the combined one- and two-tag fit, with $\epsb$ held at   the published value for each $\epsc$. White contours label fixed values of $G$; the heavy contour is break-even, $G=1$. The red curve is the operating curve of the current GN2 tagger at the jet momenta relevant to this mass. The shaded wedge at upper left is excluded by $\epsl<\epsc$. Top, $Z'\rightarrow c\bar{c}$; bottom, $T\rightarrow cg$.}
  \label{fig:contour}
\end{figure}

These sensitivities depend on the performance of the GN2 tagger, and already
improvements exist~\cite{ATLASGN3}.  To demonstrate the performance of future taggers, we compute $G$ over the plane of $\epsc$
against $\epsl$, holding $\epsb$ at the published value for each $\epsc$, and
overlay the operating curve of the current tagger; see Fig.~\ref{fig:contour}.
The contours run diagonally across the plane, so the gain is set by the pairing
of efficiency and mis-tag rate rather than by either alone.    Doubling the
sensitivity of the untagged search requires a light-jet mis-tag rate of $0.016$
at $\epsc=0.3$, compared to $0.083$ for the current tagger, a factor of $5$; at $\epsc=0.6$ the
requirement is $0.12$ compared to $0.15$, a factor of $1.3$.

% =====================================================================
\section{Conclusions}
\label{sec:conc}

Charm tagging is improving rapidly, making a dedicated search of the charm-tagged dijet mass spectrum potentially competitive with the untagged flavor-blind spectrum.

 We have predicted the background level of a dedicated charm-tagged dijet search, and estimated the expected sensitivity relative to the untagged dijet search.

 Our study finds that a narrow charm-tagged resonance such as $Z'\to c\bar c$ is accessible
        with modestly better sensitivity than the untagged search. There is also a smaller sensitivity gain to the single-charm-leg resonance.

    The perfect-tagger ceiling on the relative gain is set by the irreducible flavor
      composition of the background together with the charm content of the
      signal, and is therefore different for the two benchmarks. A perfect
      cone-label tagger would reach $G \approx 8$ for $Z'\to c\bar c$. For $T\to cg$, which populates the one-tag category against a
      background an order of magnitude larger, the ceiling is $G \approx 2.2$.
      Both ceilings are far above what current performance delivers. This suggests that there remains substantial room for enhanced sensitivity of a charm-tagged search if charm tagging improves further, as described in Fig.~\ref{fig:contour}.   
      
      The gain is largest at the loosest working point, at $\epsilon_c=0.6$ a factor of $1.3$
reduction in the light-jet mis-tag rate would double the sensitivity of the
untagged search.      Tagger development aimed at dijet resonances should therefore target the
high-efficiency working points.
      
      Our study applies binary thresholds per jet, which is less powerful than directly using the tagger's continuous output.

            The principal limitation to our study is the lack of published $\pt$-dependent charm efficiency, requiring some assumptions, and a lack of published flavor fractions at $\sqrt{s}=13$ TeV, requiring  extrapolation via simulation. Our conclusions are quantitatively sensitive to these assumptions, but survive under the variations we explore.

\begin{acknowledgments}
DW is funded by the DOE Office of Science.  DW acknowledges the use of AI tools to develop  code.  DW thanks Max Fieg, Felix Yu, David Shih, Jake Rudolph and Jesse Thaler for helpful conversations.
\end{acknowledgments}

\appendix

\clearpage

\section{Validation of the statistical treatment}
\label{app:stats}

\begin{figure}
    \centering
    \includegraphics[width=0.95\linewidth]{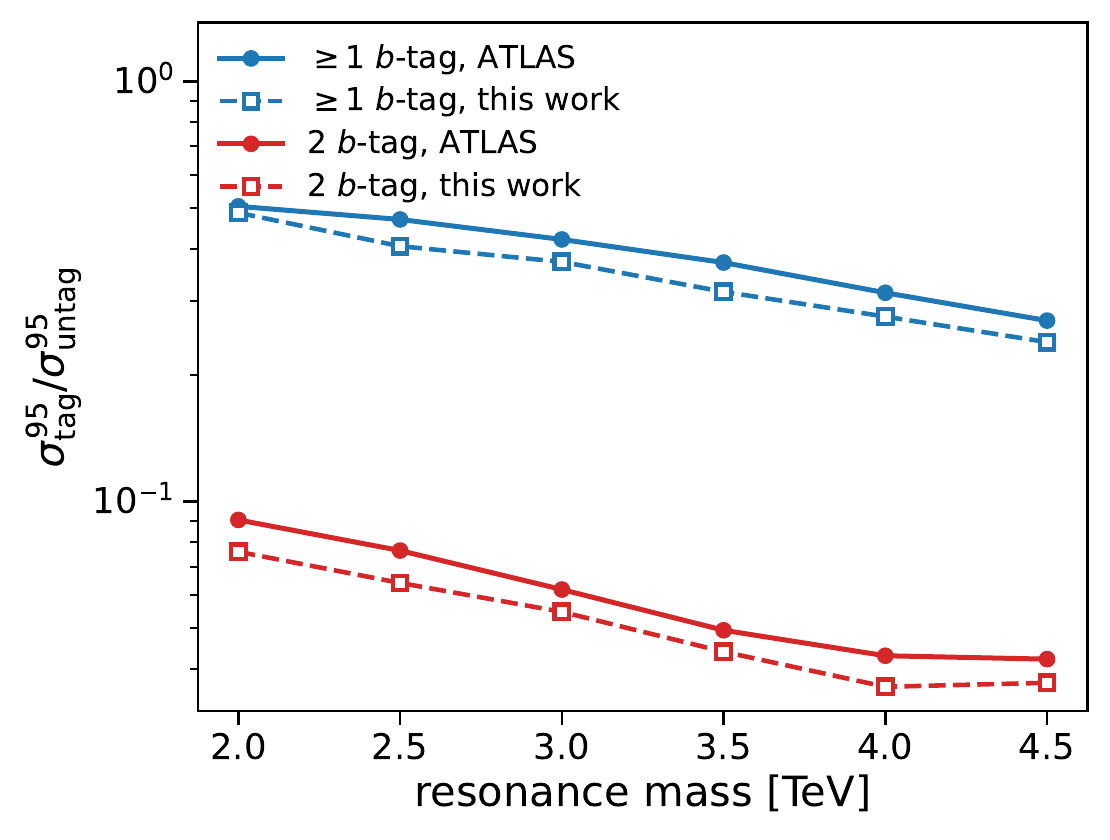}
    \caption{Validation of limit setting machinery by comparison of our calculation to results published by ATLAS using the same spectra. Shown is the ratio of   limits in the $\ge 1$- (blue) or $=2$-tagged (red) datasets to the limit in the untagged dataset, as a function of resonance mass.}
    \label{fig:limit_val}
\end{figure}

We validate our statistical limit setting treatment by deriving limits from the published ATLAS untagged and $b$-tagged datasets using our statistiacl machinery as described above and comparing our calculated limits to those from ATLAS.  We use a narrow Gaussian as the signal model, to allow direct comparison with the equivalent ATLAS limits. Our limits agree with the published results to within 10\%, despite our simplified treatment of systematic uncertainties; see Fig.~\ref{fig:limit_val}. Note that the discrepancy is roughly constant in mass and in the same direction for tagged and untagged. In this study, we only examine the ratio.

\section{Jet flavor fraction validation and uncertainty}
\label{app:flavor}

\begin{figure}[htbp]
  \centering
  \includegraphics[width=0.99\columnwidth]{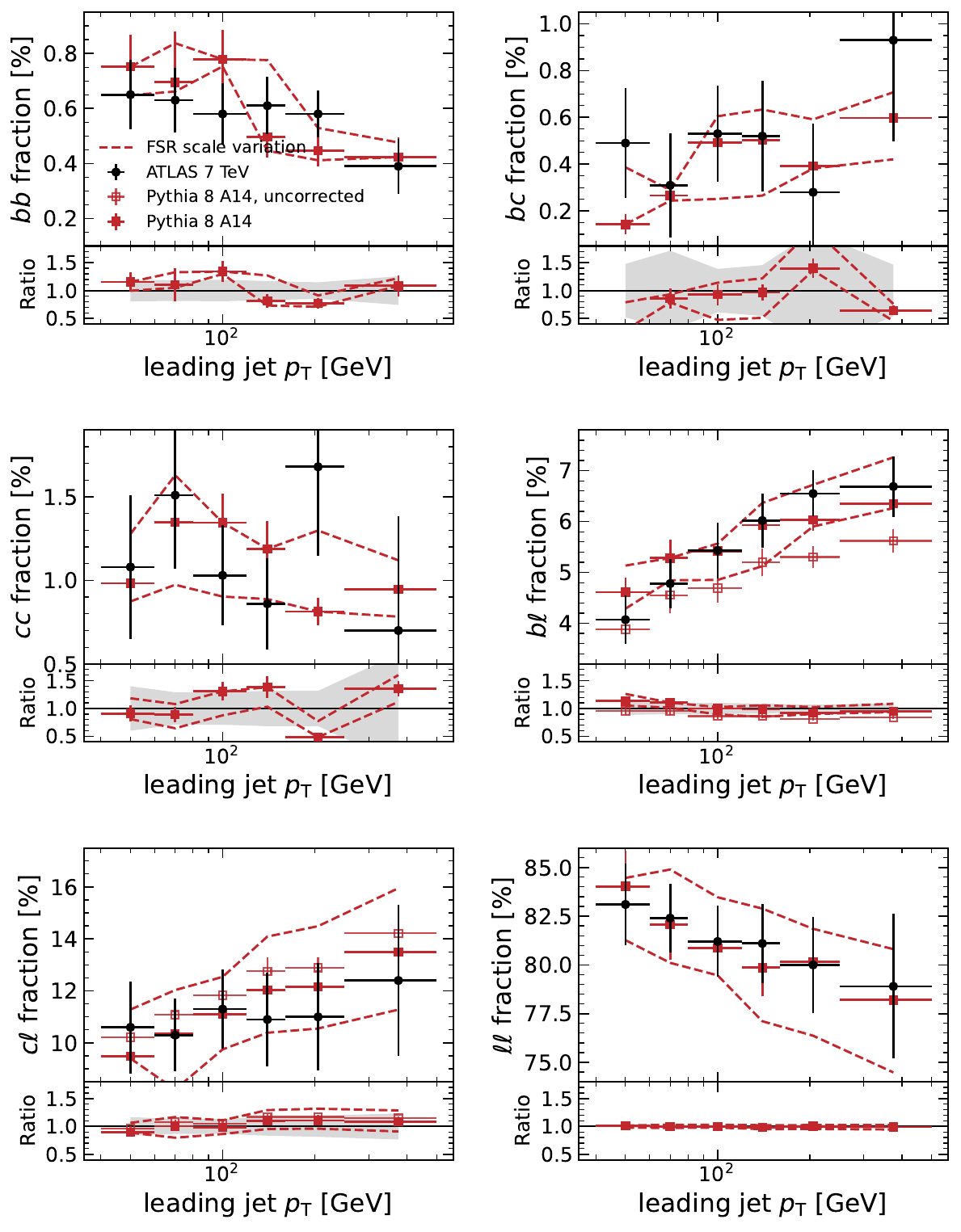}
  \caption{Dijet flavour composition at $\sqrt{s}=7$~TeV as a function of
    leading-jet $p_\mathrm{T}$, for the six pair categories. ATLAS
    data~\cite{ATLAS7TeVflavour} are compared with the nominal simulation, with a band defined by the two endpoints of the final-state radiation scale variation. The
    $b\ell$ correction is applied. Lower panels show
    the ratio to data; the shaded band is the experimental uncertainty.}
    % plots-all comes from rivet run on lxplus.cern.ch:/eos/user/w/whiteson/dijetflav
%    python3 plot_flav_band.py plots-all/ATLAS_2012_I1188891 >   --nominal local_rmf100.yoda    --band local_rmf025.yoda --band local_rmf400.yoda   --delta 0.0073 -o stage3_variations_band.pdf
  \label{fig:flavour7tev}
\end{figure}

The flavor fraction model described in the text is central to our predictions for the $b$-tagged and $c$-tagged backgrounds.

Figure~\ref{fig:flavour7tev} compares the six pair fractions ($bb$, $bc$,
$cc$, $b\ell$, $c\ell$, $\ell\ell$) calculated in simulation with 7 TeV data in six bins of leading-jet
$\pt$ from 40 to 500~GeV.  There is broad agreement except in  $f_{b\ell}$, which is low consistently; ATLAS reported the same deficiency against
\textsc{Pythia}~6.423~\cite{ATLAS7TeVflavour}.

We correct this discrepancy using the weighted mean of the data-to-simulation
difference over the six bins, $\delta = +0.0073 \pm 0.0024$ in absolute event
fraction, applied as a redistribution at fixed total heavy-flavor content: $
  f_{b\ell} \to f_{b\ell} + \delta$ and  
  $f_{c\ell} \to f_{c\ell} - \delta$.
This form acts on the pair
categories rather than rescaling the $b$-jet content per jet, which would
move $f_{bb}$ by the square of the same factor and spoil the agreement 
there. It is applied to each entry in the joint distribution of flavor
pair and jet $\pt$, scaling the $b\ell$ and $c\ell$ entries in each $\mjj$ bin
 so that the $\pt$
dependence within each category is preserved. The total yield is unchanged.
No $\pt$ dependence is applied to $\delta$ itself, as the fitted slope of the
data-to-simulation ratio, $+0.088 \pm 0.077$, is consistent with zero.
%The correction $\delta$ is derived once from the nominal sample and applied unchanged to every shower variation, so that the variations retain their sensitivity to $f_{b\ell}$. 
The uncertainty on $\delta$ is propagated as a separate nuisance
parameter.

Describing the jet flavor fractions and $\pt$ distributions within each mass bin using simulation introduces uncertainty where the simulation is unconstrained by or fails to describe collider data. For example, gluon splitting is a major source of heavy-flavor jets at high transverse momentum~\cite{ATLASbbdijet}, but the fragmentation of
        high-energy gluons at small opening angles is largely unconstrained by
        existing measurements, with significant differences observed between
        data and parton-shower predictions where it has been
        measured~\cite{ATLASgbb}.  Of course, a mismodeled flavor composition has an impact on the predicted tagged mass spectrum.  We establish an uncertainty band by varying the dominant factor, the final-state radiation renormalisation scale,
        \texttt{TimeShower:renormMultFac} $\in [0.25,4.0]$, corresponding to
        $\mu_R$ varied by a factor of two, which produces an envelope following the standard convention, but which also brackets the data  at 7 TeV, after correcting $f_{b\ell}.$   The resulting envelope changes $f_c = f_{cc} + f_{cb} + f_{c\ell}$ by $^{+22\%}_{-14\%}$ at
$\mjj \approx 1.5$~TeV, widening to $^{+28\%}_{-19\%}$ above 4~TeV relative to
nominal. 
%The band is evaluated in each $\mjj$ bin rather than quoted as a single number, since it grows with mass.

%        The resulting envelope changes $f_c$ by $^{+21\%}_{-12\%}$ at
%$\mjj \approx 1$~TeV, widening to $^{+34\%}_{-16\%}$ at $\mjj \approx 5$~TeV
%relative to nominal. The band is evaluated in each $\mjj$ bin rather than
%quoted as a single number, since it grows with mass.

The parton distribution function is varied by generating with the A14 tunes
refitted against CTEQ6L1 and MSTW2008LO. The envelope on $f_c$ is
$^{+2\%}_{-7\%}$ at $\mjj \approx 1.5$~TeV, growing to $^{+2\%}_{-13\%}$ above
4~TeV. It is strongly asymmetric, giving  lower charm
fractions than the nominal PDF over most of the mass range.

Additional shower variations were explored and found to be subdominant.  Several options for the gluon-splitting kernel, \texttt{TimeShower:weightGluonToQuark}, were compared with the default, including one that disables the suppression of high-mass $q\bar{q}$ pairs included in the nominal model, which shifts $f_c$ by $+40\%$ and
        $f_{cc}$ by $+82\%$ and is excluded by the 7~TeV data. Other gluon-splitting variants, such as a massless splitting kernel or an additional quark mass term shift $f_c$ by $-2.8\%$ and $+5.7\%$ 
        and lie inside the scale band at both $m_{jj}\sim2$~TeV and
        $m_{jj}>4$~TeV, and so are not included in the uncertainty band.
        The initial-state radiation renormalization scale,
\texttt{SpaceShower:renormMultFac}, was varied independently over the same
range and changes $f_c$ by less than $1\%$ and $f_{cc}$ by less than $3\%$;
it is not included in the band. The parameter that controls charm fragmentation was varied by
        $\pm50\%$ but changed $f_c$  by less than $1\%$; this uncertainty is not included in the band.

\begin{figure}[htbp]
  \centering
  \includegraphics[width=0.99\columnwidth]{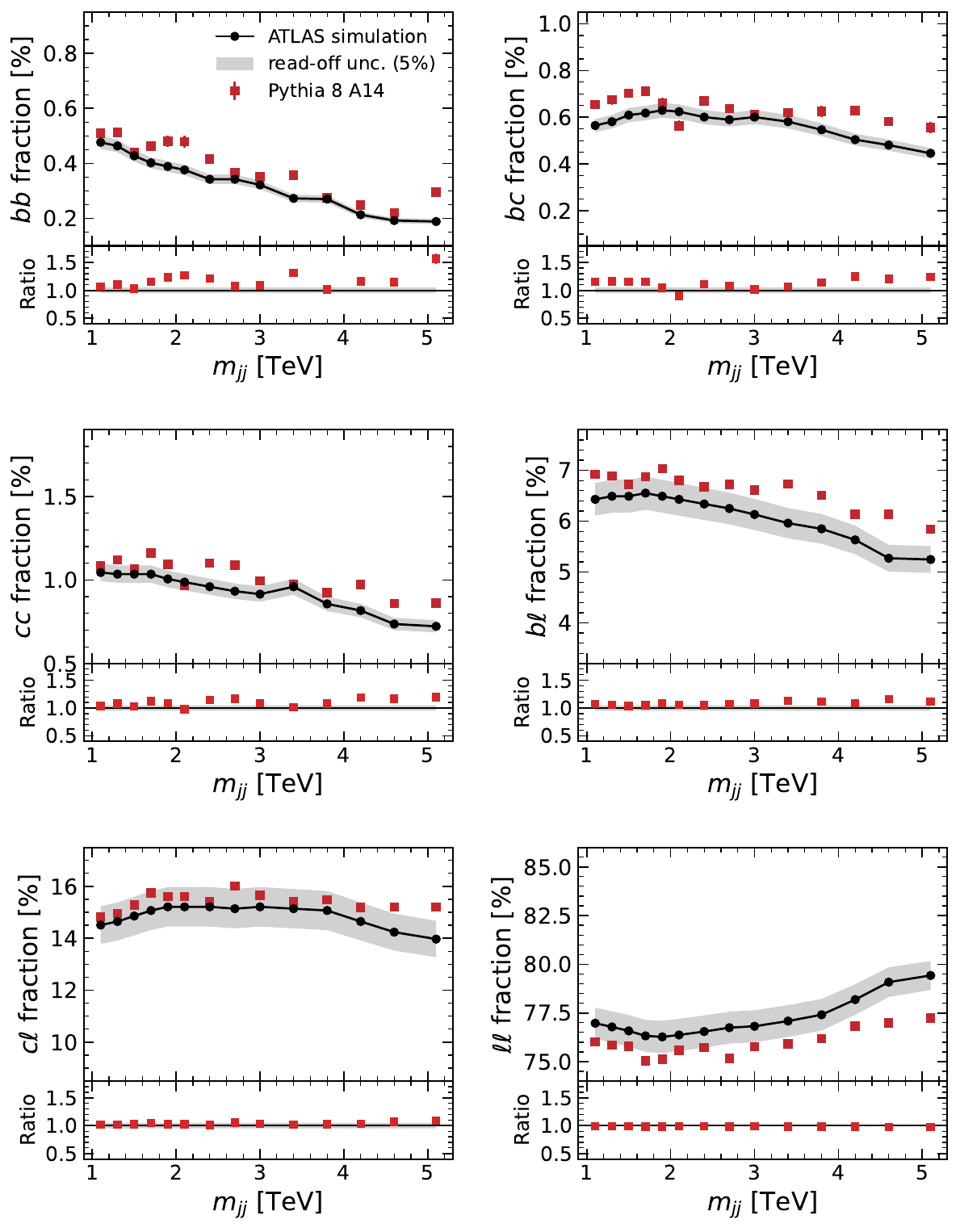}
  \caption{Inclusive dijet flavour composition at $\sqrt{s}=13$~TeV as a
    function of $m_{jj}$, for the six pair categories, compared with ATLAS
    simulation~\cite{ATLASCONF2016060}. The shaded band is the uncertainty
    assigned to the digitization of the reference. Lower panels show the ratio
    to the reference.}
  \label{fig:flavour13tev}
  % bash run_charmfrac.sh
    % python stage5.py --merge --outdir merged
%    python3 compare_conf.py merged/merged_r1.0.incl   --ref ATLAS_flavor_fractions_vs_mjj_digitized.txt -o conf_fractions.pdf
\end{figure}

        While no flavor composition data exists at $\sqrt{s}=13$ TeV, ATLAS published their tuned simulation \cite{ATLASCONF2016060} in 14 bins of $m_{jj}$ from
        $1.1$ to $5.1$~TeV; see Fig.~\ref{fig:flavour13tev}. Those flavor fractions agree with those in our sample of simulated $\sqrt{s}=13$ TeV collisions, to within uncertainties.

\begin{figure}[htbp]
  \centering
  \includegraphics[width=0.99\columnwidth]{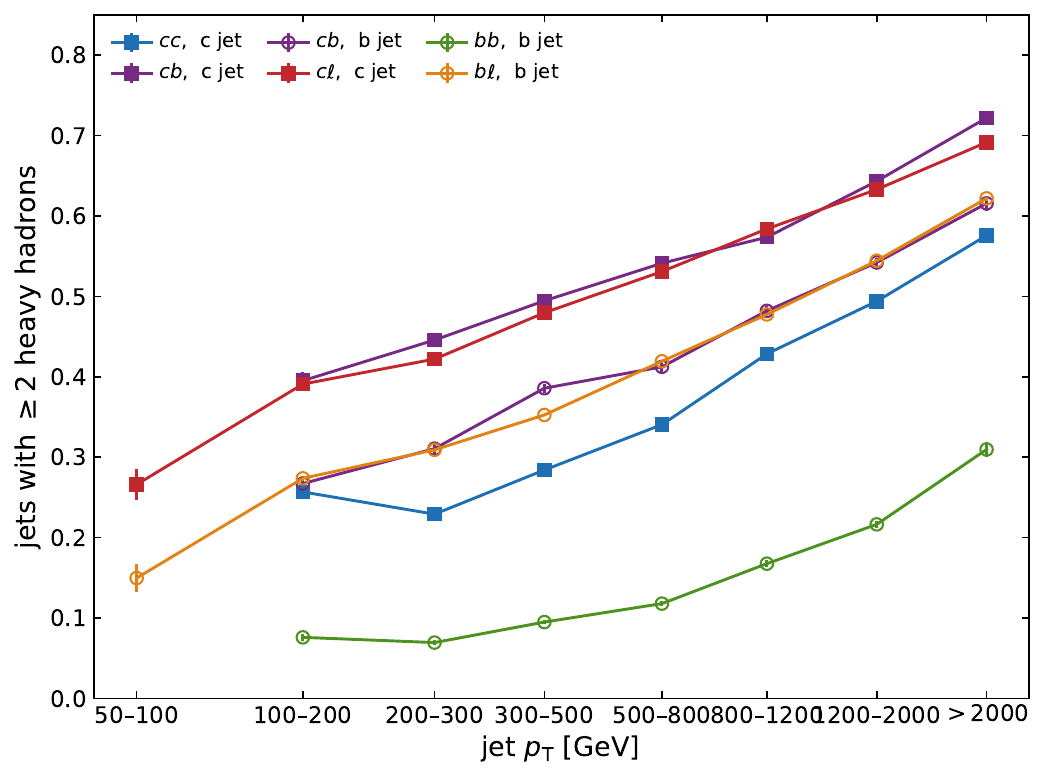}
  \caption{Fraction $f_{\rm{multi}}$ of flavor-labelled jets containing two or more
    weakly decaying heavy hadrons as a function of jet $\pt$, for each dijet flavor pair
    category. Filled markers denote charm jets and open markers bottom jets;
    color denotes the pair category. 
    %The $bb$ category, dominated by flavor    creation, is well separated from the light-paired categories, in which the    labelled jet arises largely from a gluon splitting contained within it.
    }
  \label{fig:multi}
\end{figure}

Application of the tagging efficiencies to the flavor fractions accounts for the dependence of the tagging on the flavor. However, tagging efficiency for a single flavor may also depend on whether the heavy flavor is due to a single hard quark, which gives one heavy hadron, or due to gluon splitting to small opening angles, which carries the same flavor label, but produces two heavy hadrons and a different track and vertex topology. To distinguish the sub-populations of heavy flavor, we 
count the number of weakly decaying heavy hadrons
satisfying the same requirements used to assign the label, $\pt>5$~GeV within
$\Delta R<0.3$ of the jet axis, and record the fraction $f_{\rm{multi}}$ of jets with two
or more. Figure~\ref{fig:multi} shows $f_{\rm{multi}}$
against jet $\pt$ for each pair category. The charm categories have larger $f_{\rm{multi}}$ than bottom ones,  as expected from the smaller mass
suppression of $g\to c\bar{c}$. All categories rise with jet $\pt$ as the
splittings become more collimated and both hadrons fall inside the cone. Varying the final-state shower
renormalization scale by a factor of two raises $f_{\rm{multi}}$ coherently by $9\%$,
confirming that it is driven by gluon splitting rather than by the hard process.

The description of flavor fractions and jet $\pt$ distributions in the dijet spectrum from a single generator is of course limited. Even though the envelope described covers the data variations at 7 TeV, this relies on six one-dimensional projections, which cannot capture the full extent of the high-dimensional nature of the physics, and does not guarantee coverage.  In addition, the data are limited to jets with $\pt < 500$ GeV, and in collisions at 7 TeV, and extrapolation to higher $\pt$ jets and higher center of mass relies on the single shower model and is unconstrained by data. The comparison at 13 TeV validates only our simulation against ATLAS, not against the data.   In addition, bin migration from flavor-dependent energy losses due to decays to neutrinos are not explicitly accounted for but estimated to be small.

\section{Validation in $b$-tagging}
\label{app:btag}

The robustness of predictions of the tagged spectrum, and the dependence of the tagging efficiency on the number of heavy-flavor hadrons can be studied using $b$ tags, where the observed untagged, 1- and 2-$b$-tagged spectra are available~\cite{ATLASdijet2019}. 

First, we construct a model of the $b$-tagging efficiency for $b$-jets, $c$-jets and light-flavor jets.

We estimate the DL1r tagging performance from the available information, giving 77\% $b$-jet efficiency for the DL1r~\cite{ATLASftag2022} algorithm. While more performant and better-documented taggers exist, the validation requires estimating the performance of the tagger applied to the data.  

\begin{figure}[t]
  \centering
  \includegraphics[width=\columnwidth]{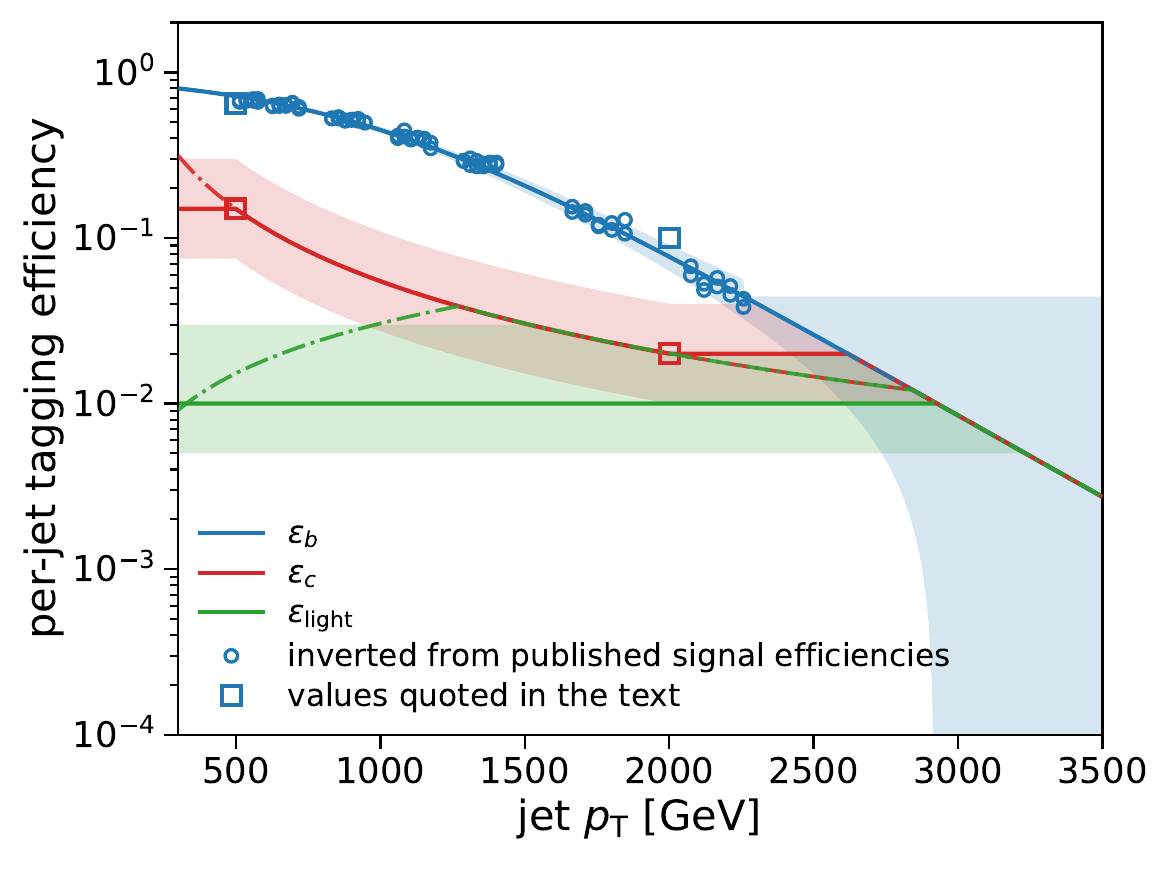}
  \caption{$b$-tagging efficiencies versus jet $\pt$. Bands show the assigned uncertainties. }
  \label{fig:btageff}
\end{figure}

 ATLAS reports event-level tagging efficiencies
for a simulated $Z'\to b\bar b$ signal as a function of resonance mass, in
both the one-tag and two-tag categories. Inverting those under the assumption
that the two jets tag independently gives a per-jet $\epsb$, and the two
categories yield consistent values, which are mapped from $\mjj$ to jet $\pt$ using the approximation 
 $\langle\pt\rangle \simeq 0.456\,\mjj$, measured in the same simulation. We fit a logistic function which matches the inversion values as well as two quoted endpoint values to better than $0.02$; see Fig.~\ref{fig:btageff}.

The tagging efficiency within a flavor category may be sensitive to the number of heavy hadrons present in each jet~\cite{CMS:2025kje}. The quoted ATLAS efficiency is in $Z'$  events, which are dominated by single-hadron jets, but the background contains significantly more multiple-hadron jets, in part due to  $g\rightarrow b\bar{b}$. 

We parameterize the efficiency $\epsilon_n$ to tag a jet carrying $n$ heavy hadrons as

\begin{equation}
    \epsilon_n = 1 - (1 - \epsilon_1)^{[1+\alpha(n-1)]}
    \label{eq:alpha}
\end{equation}

\noindent
where $\alpha=1$ indicates independent tagging probabilities per hadron and non-unity values capture deviation from that simple model.  The performance of the tagger as a function of the number of hadrons is not provided, but a value of $\alpha$ is extracted from the observed spectra below. For a given $\alpha$, the values of $\epsilon_n$ can be recovered from the efficiency quoted on the calibration $Z'$ sample, and the known distribution of hadron multiplicity, which we measure in Pythia to be $1\%-2\%$ rising with mass; see Table~\ref{tab:alpha} for examples of the multi-hadron efficiency for values of $\alpha$.

\begin{table}[htbp]
  \centering
  \caption{Per-jet $b$-tagging efficiency $\epsilon_n$ for a jet containing $n$ weakly decaying $b$-hadrons, for several values of the per-extra-hadron exponent $\alpha$; see text. }
  \label{tab:alpha}
  \begin{tabular}{llccc}
    \toprule
    $\alpha$ && $\epsilon_1$ & $\epsilon_2$ & $\epsilon_3$ \\
    \midrule
    0.5 &\ \  & 0.50 & 0.65 & 0.75 \\
    1.0 &\ \  & 0.50 & 0.75 & 0.88 \\
    1.5 &\ \  & 0.50 & 0.82 & 0.94 \\
    2.0 &\ \  & 0.50 & 0.87 & 0.97 \\
    \bottomrule
  \end{tabular}
\end{table}

ATLAS~\cite{ATLASGN2} reports a charm mis-tag rate falling from $0.15$ to $0.02$ between $\pt \simeq
500\GeV$ and $2\TeV$ and a light-jet mis-tag rate of roughly $1\%$.  We interpolate between the quoted charm endpoints and hold the light-jet rate constant, but impose the ordering $\epsl < \epsc < \epsb$  at every $\pt$. 

We adopt uncertainties due to the fit, the mass-to-$\pt$ conversion, the uncertainty on the mistag rates, and the extrapolation of the efficiency outside the fitted range, taken as the span between
continuing the logistic and freezing it at the boundary value.  

\begin{figure*}[t]
  \centering
  \includegraphics[width=0.48\textwidth]{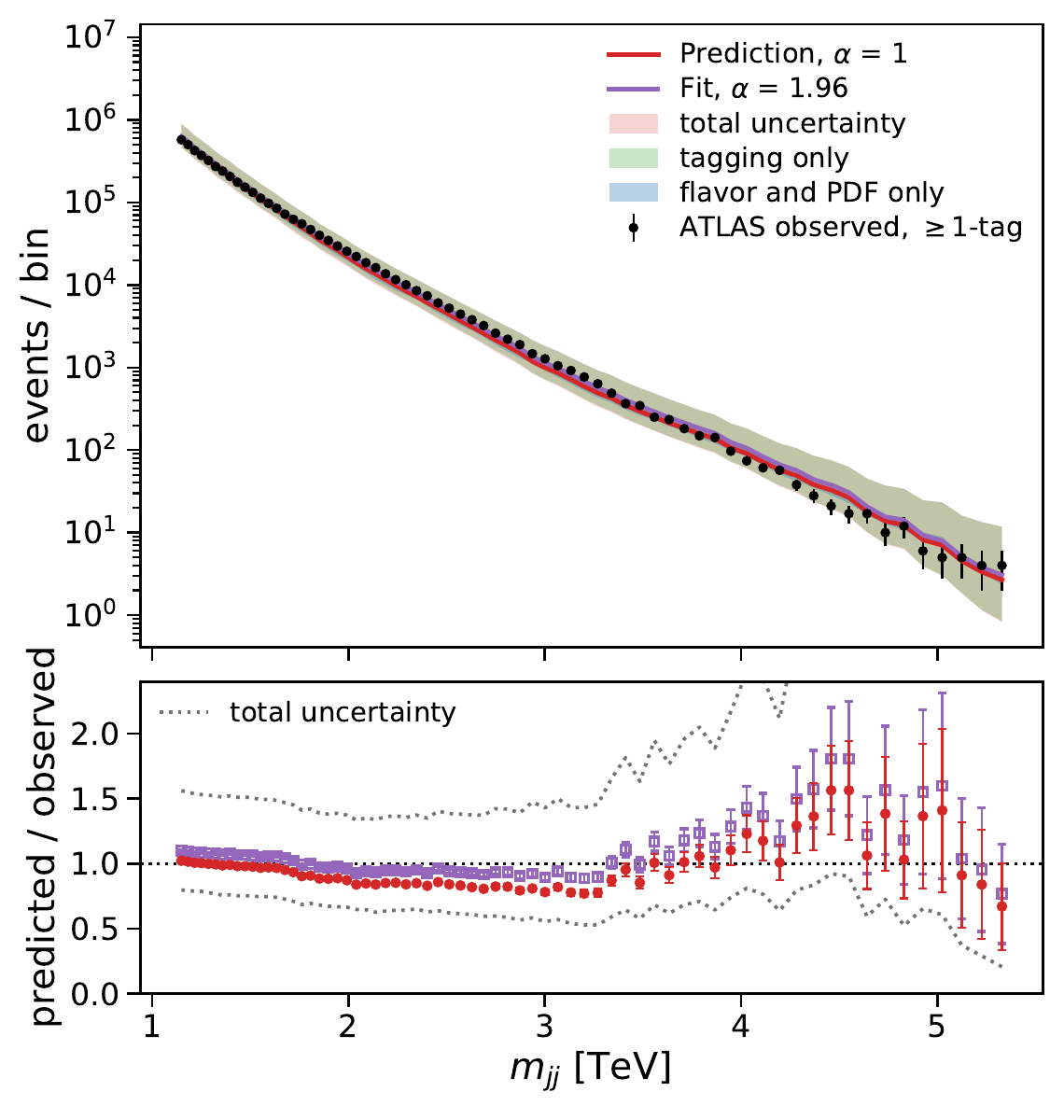}
    \includegraphics[width=0.48\textwidth]{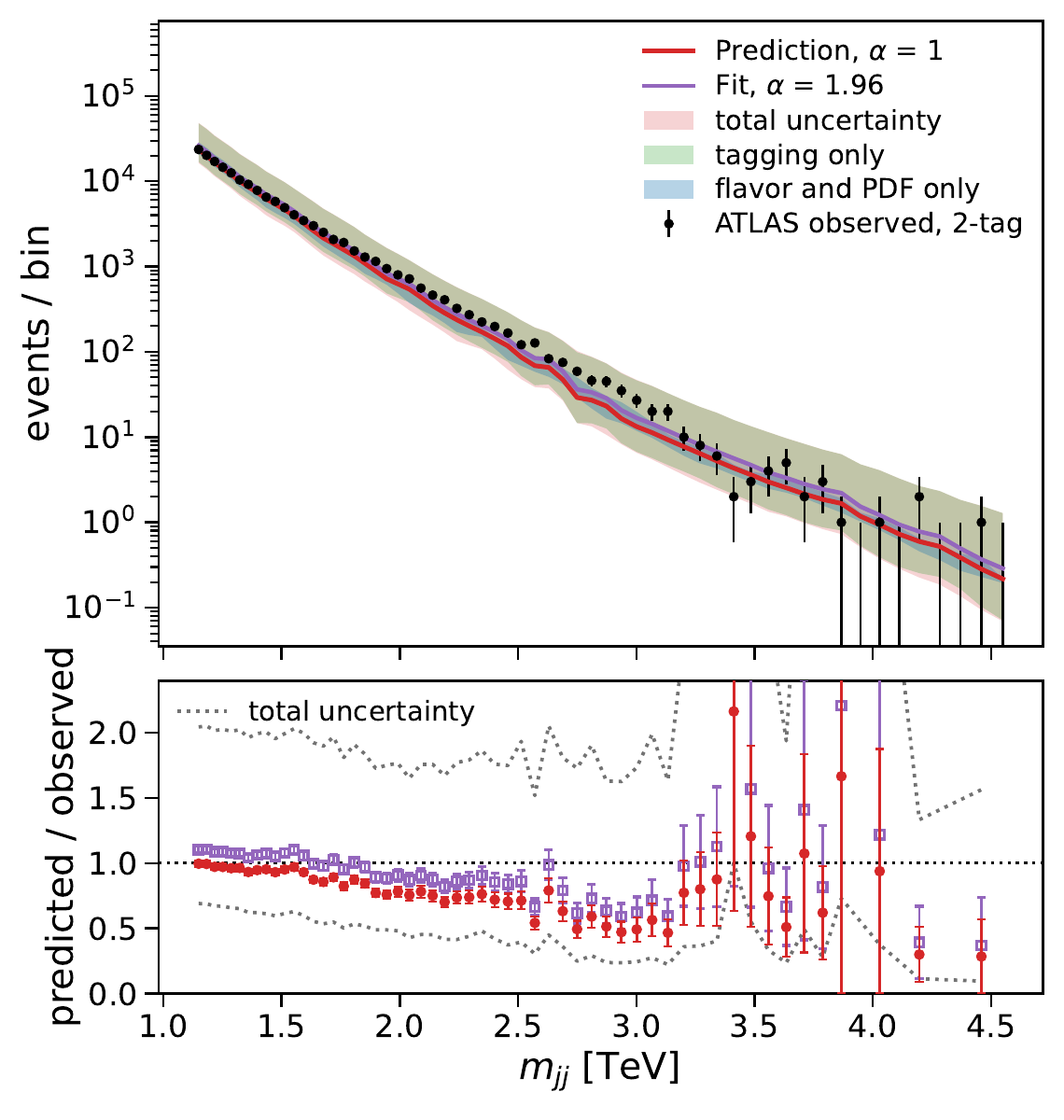}
  \caption{Predicted and observed $b$-tagged dijet spectrum using the multi-hadron-efficiency factor alpha at nominal value of $\alpha=1$ as well as the fitted value of $\alpha=1.96$, both of which are consistent with the data within uncertainties. Left: 1 tag inclusive; right, 2 tags exclusive. The bottom pane shows the ratio of the prediction to observation. }
  \label{fig:fwd}
\end{figure*}

We predict the $b$-tagged spectrum using our tagging model, see Fig.~\ref{fig:fwd}, assuming $\alpha=1$.  This prediction is below the observation but within the uncertainties, dominated by the tagging.  A fit to the observed spectra returns $\alpha=1.96$, which  slightly lowers the efficiency for single-hadron jets and raises the multiple-hadron efficiency. The simultaneous use of the one- and two-tag spectra provides sensitivity to this topology dependence because multi-hadron jets modify the two-tag probability differently from the one-tag probability. As the data accommodate both $\alpha=1$ and $\alpha=1.96$, we adopt the larger value to be conservative but consider the full range. 

No published reference is available for the tagging efficiency as a function of heavy-hadron multiplicity within a jet, but ATLAS has measured
$g\to b\bar{b}$ at small opening angles and finds it both abundant and
imperfectly described by parton showers~\cite{ATLASgbb}, and sees enhanced $b$-tagging in jets with significant gluon splitting to heavy flavor~\cite{ATLASbdijet2018}.

\section{Dependence on the charm-tagging model}
\label{app:scenarios}

The sensitivity to charmed resonances depends on the charm tagger performance as a function of jet $\pt$. However, there is limited information available about the full efficiency curves.  ATLAS has provided performance metrics in two $\pt$ regimes, [20,250] and [250,6000] GeV, which describe the degradation: at $\epsc = 0.30$ the light rejection falls from $32$ to $13.5$ and the $b$ rejection from $11$ to $5.7$.  However, that leaves unspecified  where in $\pt$ that degradation occurs,  and how it
        continues.

In the study above, we assumed a certain $\pt$ dependence. In this Appendix, we consider several scenarios to bracket our nominal choice:

    \begin{itemize}
      \item {A, early}: degradation complete by $400\GeV$.  
      \item {B, late}: little change until $\sim1\TeV$, then steep.
      \item {C, gradual}: log-uniform between anchors; 
            nominal.
      \item {D, accelerating}: as C, with the slope steepening.
      \item {E, anchor down}: high-$\pt$ anchor
            moved down.
      \item {F, anchor up}: high-$\pt$ anchor moved up.
    \end{itemize}

\begin{figure}[t]
  \centering
  \includegraphics[width=\columnwidth]{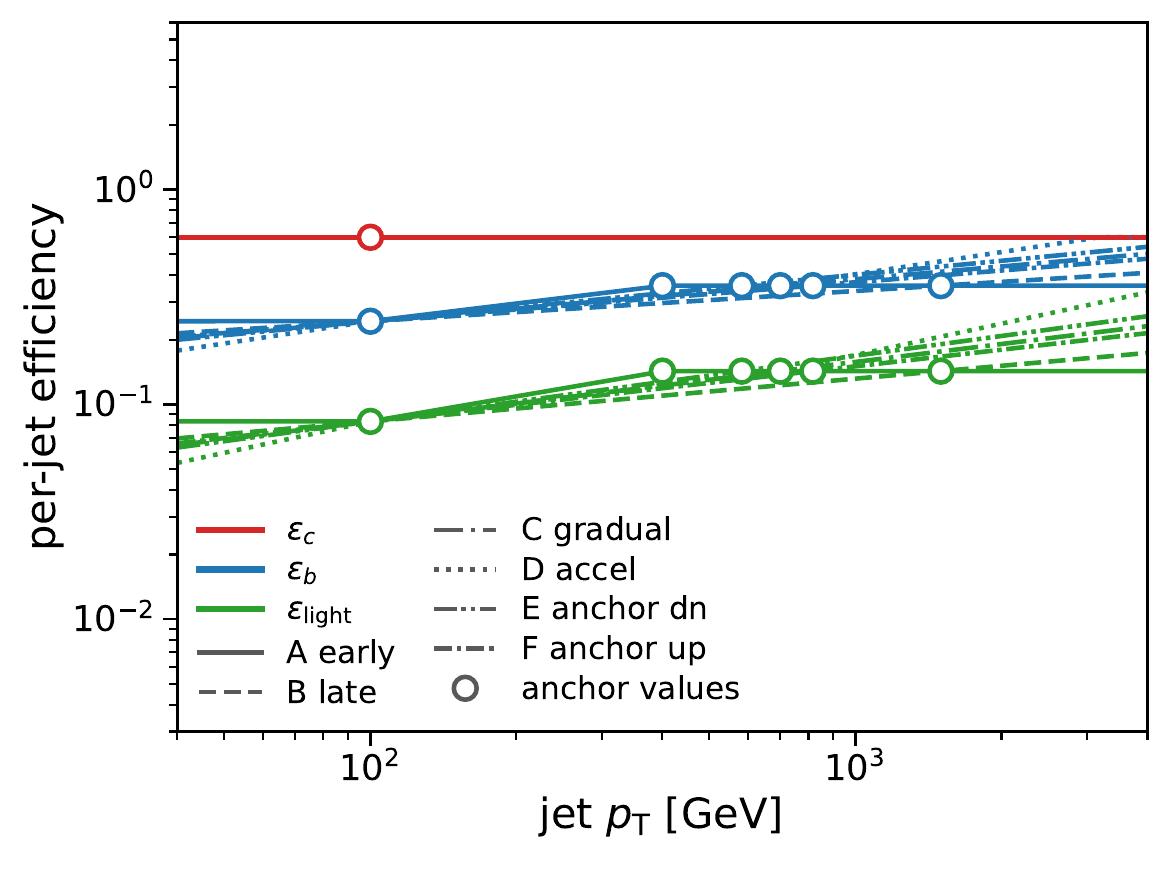}
  \caption{Six charm-tagging efficiency scenarios, drawn as per-jet efficiency versus jet $p_{\mathrm{T}}$: color denotes the flavor and line style the scenario. Open markers are the published anchor values. All scenarios share the low-$p_{\mathrm{T}}$ anchor; the high-$p_{\mathrm{T}}$ value is a single number averaged over a broad sample, so each scenario attaches it at a
different momentum.} 
  \label{fig:scenarios}
\end{figure}

Scenarios A through D each change both where the degradation happens and
how it continues beyond the anchor. To contrast, E and F hold the shape at the nominal and
move only the anchor, an unmeasured property of the
calibration sample: the published high-$\pt$ curve is a single number
averaged over jets from $250$ to $6000\GeV$, and the $\pt$ distribution of
those jets is not published, so the momentum that number describes is
uncertain by roughly the width of the sample. See Fig.~\ref{fig:scenarios}. 
These six scenarios vary the shape of the $p_{\mathrm{T}}$ dependence and and the location of the anchor, holding the anchor
efficiencies at their published values; the charm-tagging band of
Table~\ref{tab:band_zpcc} and Fig.~\ref{fig:limits} instead varies those
efficiencies within their uncertainties, so the two spreads are not
directly comparable.

An analysis by the collaborations may choose a fixed-threshold discriminant rather than the fixed-efficiency variation we have modeled here.  We do not perform a separate study for this scenario, as the performance of such a threshold reflects increasingly tighter working points at larger masses, which are already presented in the current studies.

\begin{figure}[t]
  \centering
  \includegraphics[width=\columnwidth]{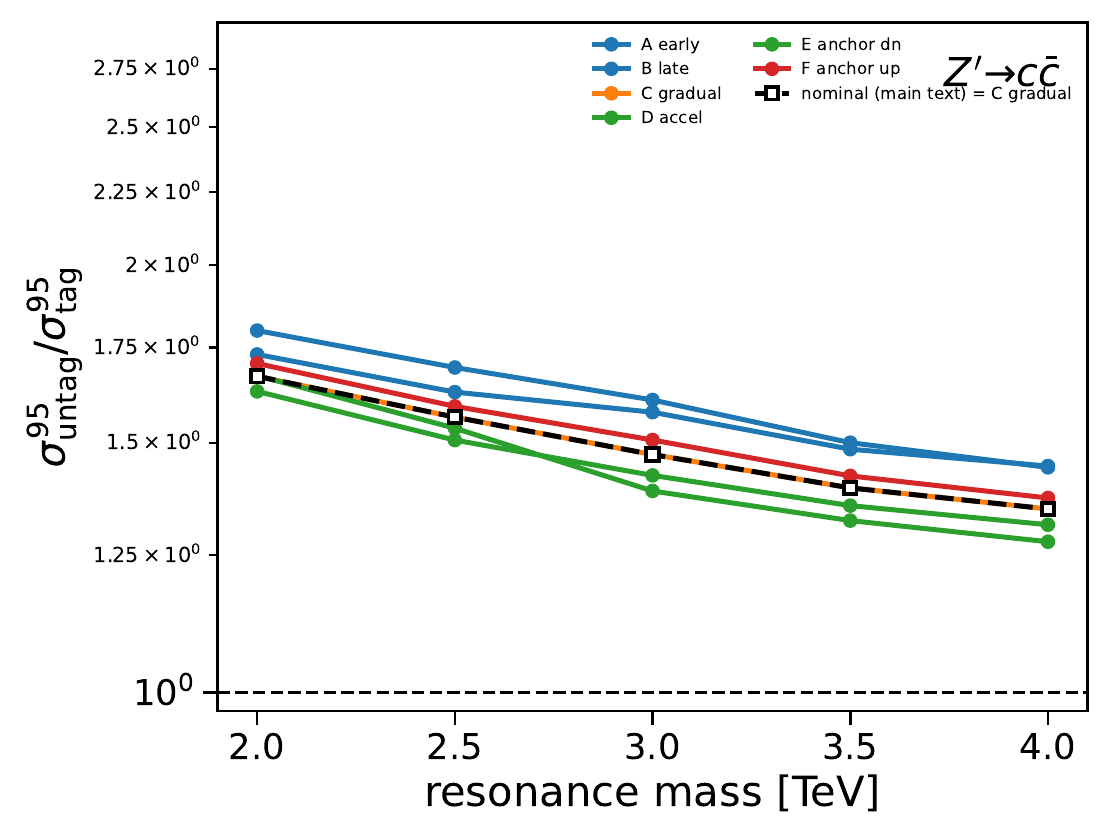}
  \caption{Sensitivity ratio $G$ under each scenario for the combined
  one- and two-tag fit at $\epsc = 0.60$. The nominal model in the text
  coincides with scenario C.}
  \label{fig:limits_scenarios}
\end{figure}

Expected gain relative to the untagged sample is shown in Fig.~\ref{fig:limits_scenarios} for each scenario.   All of them give $G > 1$ across the mass range at this working
        point.  The spread across the qualitatively different
        hypotheses is smaller than the difference between working points. A charm tag improves the limit at loose working
        points for every scenario considered, including the pessimistic
        one.

% ---------------------------------------------------------------------------
% Appendix: jet flavour labelling and its infrared sensitivity
%
% Numbers are from the nominal IFN production (alpha = 2, omega = 1, net
% summation; 7.8M events over five pThat slices) and are reproducible from the
% .migration and .incl files in that output directory.  Citation keys are
% placeholders.  Full migration tables, both fiducials and per jet pT bin, are
% in the supplementary material.
%
% This appendix reports a cross-check.  It does not change the background
% prediction, the efficiencies, the limits or the shower band, and should not
% be rewritten so that it appears to.
% ---------------------------------------------------------------------------

\section{Jet flavor labeling and its infrared sensitivity}
\label{app:ifn}

The ATLAS jet labeling convention~\cite{ATLASGN2} follows weakly-decaying heavy-flavor hadrons near the jet axis: $b$ if a $b$-hadron with $p_{\mathrm{T}} > 5$~GeV lies within
$\Delta R < 0.3$, $c$ if a $c$-hadron does and no $b$-hadron does, light
otherwise. However, this prescription is not infrared- and collinear- (IRC-) safe: a single soft
$g \to c\bar{c}$ anywhere in the jet affects the label, making the labeled fractions uncalculable  in perturbation theory, with sensitivity growing
with $p_{\mathrm{T}}$, where gluon splitting dominates.

The study above retains the cone-labeling scheme despite its lack of IRC safety, as the tagger efficiencies are only reported for this labeling. In addition, the IRC-unsafe quantities are the features that drive the tagger: a collimated $c\bar{c}$ pair produces real displaced vertices to tag, regardless of its flavor label.  

However, claims about the perturbative flavor composition of the QCD background should be made using IRC-safe labeling. To quantify the size of the effect,  the composition was
re-measured using interleaved flavor neutralization
(IFN)~\cite{caola:ifn}, which assigns flavor in an IRC-safe way while leaving
the anti-$k_t$ kinematics  unchanged, neutralizing soft $q\bar{q}$ pairs
before their flavor can reach a harder jet. We use the authors'
implementation~\cite{caola:ifnplugin}, with parameters $\alpha = 2$,
$\omega = 1$ and net flavor summation, applied to the same events  with no $p_{\mathrm{T}}$
threshold.

\begin{figure}[htbp]
  \centering
  \includegraphics[width=0.9\columnwidth]{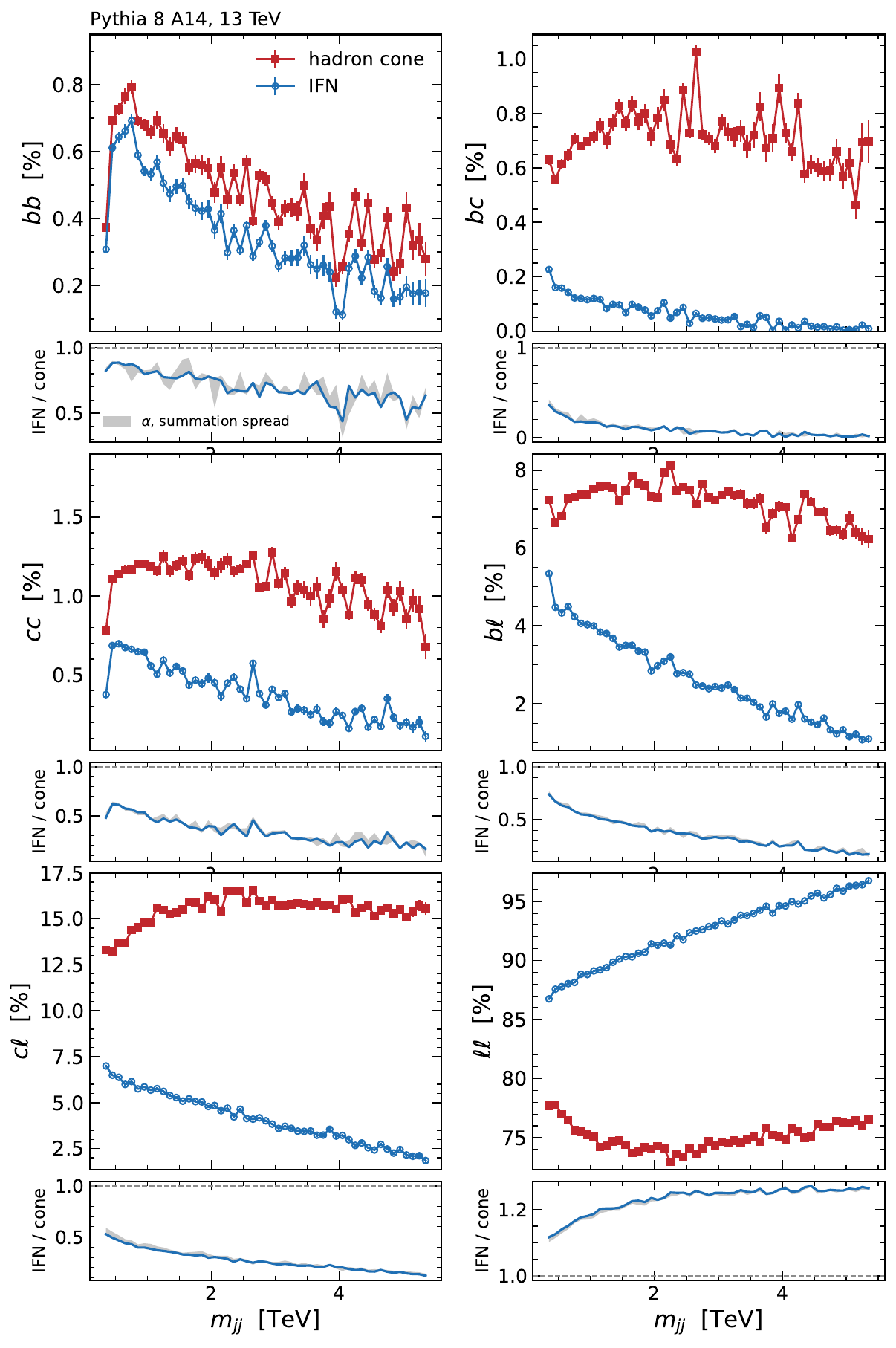}
  \caption{Flavor composition of the simulated QCD dijet background versus
    $m_{jj}$, under the hadron-cone labeling used in this analysis (filled
    squares) and under the IRC-safe IFN labeling~\cite{caola:ifn} (open circles). Lower panels give the ratio; the shaded band is the spread over
    parameters as described in the text. Point uncertainties are statistical.}
  \label{fig:ifn-composition}
\end{figure}

Figure~\ref{fig:ifn-composition} shows the labels under both definitions and Table~\ref{tab:ifn-migration} shows the migration. Roughly half the jets the cone label calls charm at low mass are light under an IRC-safe definition; the fraction grows with mass.   But changes to the labeling do not affect the prediction for the tagged spectrum, only what labels are applied.

\begin{table}[htbp]
  \centering
  \caption{Migration between the cone and IFN labels, per jet, tagged fiducial
    region; entries are fractions of the cone row.}
  \label{tab:ifn-migration}
  \begin{tabular}{lcccccc}
    \hline\hline
    Mass & \multicolumn{2}{c}{charm}&\ \ & \multicolumn{3}{c}{bottom}\\
    $m_{jj}$ [TeV] & $c \to c$ & $c \to \ell$ & &$b \to b$ & $b \to c$
                   & $b \to \ell$ \\
    \hline
    $0.3 - 0.6$ & 0.497 & 0.503 && 0.682 & 0.0033 & 0.314 \\
    $1.0 - 2.0$ & 0.373 & 0.627 && 0.521 & 0.0067 & 0.472 \\
    $4.0 - 5.4$ & 0.172 & 0.828 && 0.249 & 0.0089 & 0.742 \\
    \hline\hline
  \end{tabular}
\end{table}

The result is insensitive to the algorithm's parameters: over
$\alpha \in \{1,2\}$ with $\omega = 3 - \alpha$ and both flavor summation
schemes, the overall charm migration varies by $1\%$, drawn as the band
in Fig.~\ref{fig:ifn-composition}. Repeating the measurement on the
pre-hadronization partons of the same events changes the fractions by no more
than $2\%$ relative in any category~\cite{behring:comparison}.

\bibliography{refs}

\end{document}